\documentclass[
nofootinbib,
noeprint,
amsmath,amssymb,
aps,
nobibnotes,
twocolumn,
floatfix,
]{revtex4-2}

\usepackage[utf8]{inputenc}
\usepackage{graphicx}
\usepackage{amsmath}
\usepackage{amssymb}
\usepackage{bm}
\usepackage{txfonts}
\usepackage{multirow}
\usepackage{mathptmx} 
\usepackage{siunitx}
\usepackage{xargs}
\usepackage{printlen}
\usepackage{booktabs}
\usepackage[pdftex,dvipsnames]{xcolor}
\usepackage[colorlinks=true, urlcolor=blue, linkcolor=blue, citecolor=blue]{hyperref}
\usepackage{xcolor}
\usepackage{braket}
\usepackage{soul}
\usepackage{verbatim}

\begin{document}

\title{Generating arbitrary superpositions of nonclassical quantum harmonic oscillator states}

\date{\today}
\author{S. Saner, O. B\u{a}z\u{a}van, D. J. Webb, G. Araneda, D. M. Lucas, C. J. Ballance, R. Srinivas \\
\normalsize{Department of Physics, University of Oxford, Clarendon Laboratory, \\
Parks Road, Oxford OX1 3PU, United Kingdom \\Email: sebastian.saner@physics.ox.ac.uk}}

\begin{abstract}
Full coherent control and generation of superpositions of the quantum harmonic oscillator are not only of fundamental interest but are crucial for applications in quantum simulations, quantum-enhanced metrology and continuous-variable quantum computation. The extension of such superpositions to nonclassical states increases their power as a resource for such applications. Here, we create arbitrary superpositions of nonclassical and non-Gaussian states of a quantum harmonic oscillator using the motion of a trapped ion coupled to its internal spin states. We interleave spin-dependent nonlinear bosonic interactions and mid-circuit measurements of the spin that preserve the coherence of the oscillator. These techniques enable the creation of superpositions between squeezed, trisqueezed, and quadsqueezed states, which have never been demonstrated before, with independent control over the complex-valued squeezing parameter and the probability amplitude of each constituent, as well as their spatial separation. We directly observe the nonclassical nature of these states in the form of Wigner negativity following a full state reconstruction. Our methods apply to any system where a quantum harmonic oscillator is coupled to a spin.
\end{abstract}
\maketitle


One of the most fundamental and remarkable aspects of quantum mechanics is superposition, where quantum objects can exist in multiple states simultaneously. This phenomenon was famously described by Schr\"{o}dinger using the example of a cat that is both dead and alive~\cite{schrodinger1935die}. Superpositions are not just a curiosity of nature but are central to any application of quantum mechanics. In sensing~\cite{degen2017quantum}, it is the superposition of two states that respond differently to a given signal, such as a shift in frequency, that enables their use in atomic clocks~\cite{ludlow2015optical}. 
In quantum computing, the ability to be in multiple states at once increases the amount of information that can be stored and processed simultaneously and is the basis of quantum advantage~\cite{preskill2012quantum}. 

Superpositions of two-level systems can be controlled to almost arbitrary precision in many systems~\cite{leu2023fast,barends2014superconducting,yoneda2018a}. However, they ultimately only have two degrees of freedom: the relative amplitude and phase of their complex probabilities, which limits the amount of quantum information and how robustly it can be stored. Creating superpositions of a quantum harmonic oscillator, which is, in contrast, infinite-dimensional, has already led to a plethora of resource-efficient error correction protocols~\cite{gottesman2001encoding, cochrane1999macroscopically, marios2016new}, metrological applications~\cite{bild2023schroedinger}, and quantum simulations~\cite{kang2024seeking}. Nonetheless, constituents of those superpositions have been limited to two~\cite{mccormick2019quantum, kudra2022robust, matsos2023robust} or three~\cite{bimbard2010quantum} Fock states, classical states such as coherent states leading to cat states~\cite{deMatos1996even,ourjoumtsev2007generation, kienzler2016observation,heeres2017implementing, kudra2022robust}, or Gaussian states such as displaced squeezed states leading to Gottesman-Kitaev-Preskill (GKP) states~\cite{gottesman2001encoding,fluhmann2019encoding,campagne2020quantum,shunya2024logical}. 


Instead, superpositions between arbitrary squeezed states~\cite{sanders1989superposition,barbosa2000generalised,nicacio2010phase,mcconnell2022multisqueezed,ayyash2024driven}, or between non-Gaussian states such as trisqueezed states, exhibit increased Wigner negativities~\cite{albarelli2018resource, takagi2018convex} and discrete rotational symmetries which can be a resource for new sensing~\cite{drechsler2020state} and error correction schemes~\cite{grimsmo2020quantum}.
However, the lack of any strong, native interactions that create such superpositions has precluded any experimental realisation.

Here, we use a hybrid oscillator-spin system to create such superposition states~\cite{sanders1989superposition} of the harmonic oscillator; we use our recently demonstrated strong spin-dependent nonlinear interactions of the oscillator~\cite{bazavan2024squeezing} together with mid-circuit measurements of the spin. We create an arbitrary combination of generalised squeezed states~\cite{braunstein1987generalised} such as squeezed, trisqueezed, and quadsqueezed harmonic oscillator states.
These techniques not only enable arbitrary control of the complex probability amplitudes of the constituents forming the superposition, but also over the interactions that generate them. We can create constituents from interactions of the same type with different strengths and phases, or from entirely different nonlinear interactions. 
\begin{figure}[!ht]
\includegraphics[width=\linewidth]{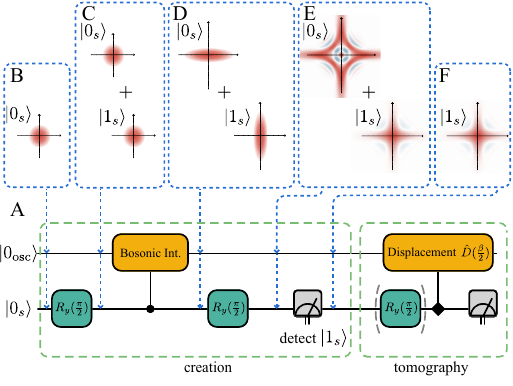}
\caption{
Quantum circuit for generating superpositions of a quantum harmonic oscillator.
We utilise a hybrid system consisting of a harmonic oscillator ($\ket{\textrm{osc}}$) coupled to a spin ($\ket{0_s}, \ket{1_s}$). As an example, we illustrate the creation of a superposition of states squeezed along two different axes (i.e. $\ket{\zeta_k}$ and $\ket{-\zeta_k}$). However, the same procedure would apply to any other spin-dependent bosonic interaction.
({\bf A}) Circuit diagram. The insets ({\bf B})-({\bf F}) illustrate the state of the hybrid system at the points of the circuit indicated. At each step, we show the Wigner function of the oscillator state and the corresponding state of the spin. 
({\bf B}) We initialise the hybrid system in $\ket{0_{osc},0_s}$. ({\bf C}) We first apply a $R_y(\pi/2)$ rotation to the spin qubit to create a superposition of $\ket{0_s}$ and $\ket{1_s}$. ({\bf D}) After applying the $\hat{\sigma}_z$ spin-conditioned (black dot) bosonic interaction (i.e.~generalised squeezing), the oscillator and spin are in an entangled superposition state. ({\bf E}) The second qubit rotation $R_y(\pi/2)$ brings the hybrid state into a superposition of the form $(\ket{\zeta_k}-\ket{-\zeta_k})\ket{0_s}$ and $(\ket{\zeta_k}+\ket{-\zeta_k})\ket{1_s}$ where $\ket{0_s}$ and $\ket{1_s}$ are entangled with the odd and even harmonic oscillator superposition, respectively (see Eq.~\ref{eq:qubit_entangled_superpos}). ({\bf F}) 
A subsequent measurement projects the state to either the odd or even superposition. We show a projective measurement outcome corresponding to $\ket{1_s}$ that heralds the even superposition. 
The creation step can be generalised (see Fig.~\ref{fig:arbitrary_superpos}) and concatenated (see Fig.~\ref{fig:concatenate_sequence}) to build up arbitrary superposition states. After the creation step(s), we perform state tomography of the final projected oscillator state by measuring the characteristic function of the state~\cite{fluhmann2020direct} and use it to reconstruct the Wigner function. To measure the real part of the characteristic function, we apply a $\hat{\sigma_y}$ conditioned displacement (black diamond) and readout the spin. By adding a $R_y(\pi/2)$ pulse before the conditional displacement, we can measure the imaginary part of the characteristic function instead.
}
\label{fig:scheme_equal}
\end{figure}
\begin{figure*}[t]
\includegraphics[]{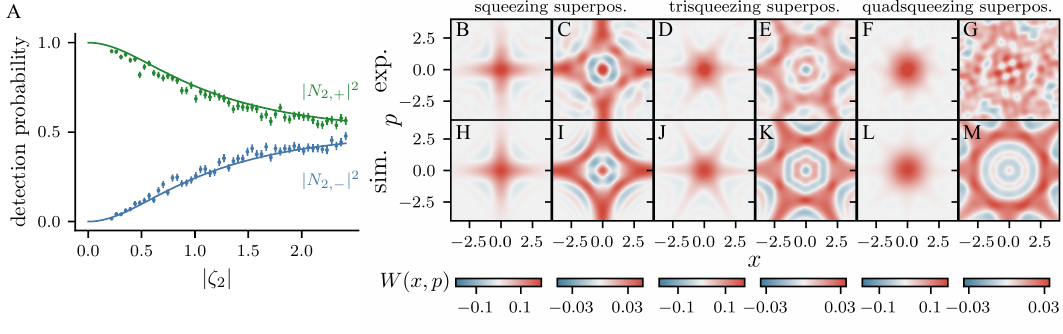}
\caption{Equal superpositions of generalised squeezed states.
({\bf A}) Probability of creating even (green dots) and odd (blue dots) squeezed superposition states $\ket{\psi_{2,\pm}} =1/(2N_{2, \pm}) \cdot ( \ket{\zeta_2}\pm \ket{-\zeta_2})$ as a function of the squeezing magnitude $|\zeta_2|$. The theoretical detection probability values $|N_\pm|^2$ predicted by the coefficients in Eq.~\eqref{eq:qubit_entangled_superpos} are displayed as solid lines. At large values of $|\zeta_2|$, the detection probabilities $|N_{2,\pm}|^2$ approach 0.5. Error bars denote the 68\% confidence interval. 
 ({\bf B}) - ({\bf G}) Experimentally reconstructed Wigner functions of even and odd superpositions $\ket{\psi_{k,\pm}} =1/(2N_{k,\pm})\cdot ( \ket{\zeta_k}\pm \ket{-\zeta_k})$ shown for squeezed, trisqueezed, and quadsqueezed superpositions. The estimated squeezing magnitudes are $|\zeta_2| \approx \{1.12(5), 1.67(7) \}$ for squeezed (B, C), $|\zeta_3| \approx \{0.74(5), 0.74(5)\}$ for trisqueezed (D, E), and $|\zeta_4| \approx\{0.059(5), 0.16(1)\}$ for quadsqueezed (F, G) superpositions.
 The experimental parameters are tabulated in the Supplement~\cite{supplementary}. For the odd states of the squeezed and quadsqueezed superpositions, we have increased the squeezing magnitude relative to the even state to highlight the interference effects. ({\bf H}) - ({\bf M}) Simulated Wigner function of panels (B) - (G) using independently measured experimental parameters.
}
\label{fig:equal_superpositions}
\end{figure*}
The hybrid system is central to our demonstration. 
Applying the spin-dependent nonlinear interaction to a superposition of spin states leads to
an entangled state of the spin and the nonclassical motional states.
A mid-circuit measurement of the spin subsequently disentangles the spin from the motion, leaving the harmonic oscillator in a superposition state. 
This method enables the repeated application of the nonlinear interaction within a single sequence, as the interaction is unitary and can act on any starting state without destroying it.

We create the spin-dependent nonlinear interaction
\begin{equation}\label{eq:int_hamiltonian}
    \hat{H}_k =\frac{\hbar\Omega_k}{2} \hat{\sigma}_z (\hat{a}^k e^{-i\phi} + (\hat{a}^\dagger)^k e^{i\phi}),
\end{equation}
by combining two noncommuting spin-dependent forces (SDF)~\cite{sutherland2021universal, bazavan2024squeezing} (see Supplement~\cite{supplementary}). 
The spin Pauli operator is defined as $\hat{\sigma}_z = \ket{1_s}\bra{1_s} - \ket{0_s}\bra{0_s}$, while $\hat{a}^\dagger$ ($\hat{a}$) describes the creation (annihilation) operator of a quantum harmonic oscillator. The order of the generalised squeezing interaction~\cite{braunstein1987generalised} is $k$; setting $k=2, 3, 4$ generates the squeezing, trisqueezing, and quadsqueezing interactions respectively. Squeezing interactions have been used extensively in metrology~\cite{caves1981quantum, burd2019quantum}, while the higher-order trisqueezing~\cite{chang2020observation, eriksson2023universal,bazavan2024squeezing} and quadsqueezing~\cite{bazavan2024squeezing} create non-Gaussian states of the harmonic oscillator~\cite{sigmazforsimplicity}.
The coupling strength of the interaction is $\Omega_k$, and $\phi$ is the phase relative to the oscillator, which also defines the generalised squeezing axis.
When applying this interaction to the initial state of the hybrid system $\ket{0_{osc},0_s}$, where $\ket{0_{osc}}$ represents the harmonic oscillator in its ground state and $\ket{0_s}$ denotes the spin configuration prepared in an eigenstate of $\hat{\sigma}_z$, we obtain a generalised squeezed state $\ket{\zeta_k, 0_s}:=\exp(-i\hat{H}_k t/\hbar)\ket{0_{osc}, 0_s}$ with squeezing magnitude $\zeta_k = \Omega_k e^{i\phi} t$.

A superposition between two generalised squeezed states $\ket{\zeta_k}$ and $\ket{-\zeta_k}$, with orthogonal squeezing axes, can be created by applying the sequence shown in Fig.~\ref{fig:scheme_equal}A. Initially, a $R_y(\pi/2)$ rotation brings the spin part of the ion into a superposition state $\ket{+_s}= (\ket{0_s}+\ket{1_s})/\sqrt{2}$ before applying the $\hat{\sigma}_z$ conditioned nonlinear coupling. After the nonlinear coupling and another $R_y(\pi/2)$ rotation have been applied, the resulting entangled state of the hybrid system takes the form

\begin{equation}\label{eq:qubit_entangled_superpos}
 \ket{\psi} = N_{k,+}\ket{\psi_{k,+}}\ket{1_s} + N_{k,-}\ket{\psi_{k,-}}\ket{0_s},
\end{equation}
where $N_{k,\pm}$ denote the probability amplitudes of the even ($\ket{\psi_{k,+}}$) and odd ($\ket{\psi_{k,-}}$) superpositions of the oscillator, which arise directly from the normalisation coefficient of ${\ket{\psi_{k,\pm}} = 1/(2N_{k,\pm})\cdot (\ket{\zeta_k} \pm \ket{-\zeta_k})}$. The normalisation coefficients depend on the overlap between $\ket{\zeta_k}$ and $\ket{-\zeta_k}$ and, consequently, on the magnitude and order of the applied nonlinear interaction.

To project into the desired oscillator superposition state, we perform a mid-circuit measurement detecting $\ket{0_s}$ or $\ket{1_s}$, which occurs with a probability of $|N_{k,\pm}|^2$. This measurement collapses the state into the corresponding superposition of the oscillator, disentangled from the spin state. After the collapse, the created superposition can be employed in further experiments, or characterised through full tomography. 
Experimentally, we can select the outcomes with the desired superposition by postselecting based on the mid-circuit measurement result.
\begin{figure*}[!ht]
\includegraphics[]{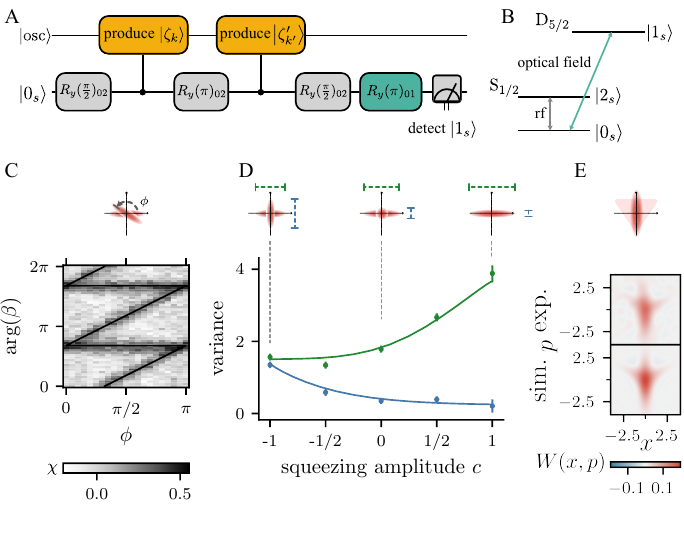}
\caption{Generation of superpositions with arbitrary constituents. ({\bf A}) The sequence used to create these superpositions is similar to Fig.~\ref{fig:scheme_equal}A, but extended to a qutrit system such that the constituents ($\ket{\zeta_k}$, $\ket{\zeta_{k'}'}$) can be produced sequentially. The indices next to the rotations $R_y()_{ij}$ and colours indicate the two spin sublevels $\ket{i} \leftrightarrow \ket{j}$ coupled. ({\bf B}) Level diagram of the three-level spin system, using both levels of the ground $S_{1/2}$ state, and one level in the excited $D_{5/2}$ state. We use radiofrequency (rf) pulses to perform rotations between $\ket{0_s}\leftrightarrow \ket{2_s}$, and optical fields for rotations between $\ket{0_s}\leftrightarrow\ket{1_s}$. The bosonic interaction is performed by optical fields and hence, only couples the $\ket{0_s}$ and $\ket{1_s}$ states. ({\bf C}) 
We first create superposition states with the same squeezing amplitude but arbitrary phase $\phi$ i.e.~$\ket{\zeta_2}+ \ket{e^{i\, 2\phi} \zeta_2}$. To verify control over $\phi$ and consequently the angle between the squeezing axes of the two constituents, we measure the value of the characteristic function $\chi(\beta)$ at a fixed magnitude $|\beta|$ but vary its complex argument $\arg(\beta)$. Maxima in $\chi$ indicate the location of squeezing axes of the two constituents.
The estimated squeezing magnitude is $|\zeta_2| = 1.12(5)$. ({\bf D}) We next vary the squeezing amplitude between the states i.e.~$\ket{\zeta_2}+ \ket{c \cdot \zeta_2}$ and measure the variance of the state along $x$ (green) and $p$ (blue) for various settings of $c$. Error bars denote the 68\% confidence interval via bootstrapping. Solid lines denote the numerically simulated value for the variance. 
({\bf E}) We show the experimentally reconstructed Wigner function of an even superposition state between a squeezed and trisqueezed state $\ket{\zeta_2}+\ket{\zeta_3}$ (upper row) as well as the simulated Wigner function (lower row). The squeezing and trisqueezing magnitudes are $\zeta_2 \approx 1.12(5)$ and $\zeta_3 \approx 0.25(2)$, respectively.
The insets show the Wigner function of each of the constituents overlapped to illustrate the expected resulting Wigner functions.
The insets do not depict the coherences between the two constituents, which are present in the real superposition states.
The experimental parameters are tabulated in the Supplement~\cite{supplementary}.}
\label{fig:arbitrary_superpos}
\end{figure*}

We experimentally demonstrate the creation of these superpositions on a trapped $^{88}$Sr$^+$ ion in a 3D Paul trap~\cite{schafer2018fastthesis,thirumalai2019high}. The harmonic oscillator is formed by the axial motion of the confined ion with frequency \SI{1.2}{\MHz}. In practice, we cannot initialise a pure vacuum state of the oscillator $\ket{0_{osc}}$ but rather a thermal state near its ground state with average occupation $\bar{n}=0.1$. The coherence time of this oscillator is limited by its heating rate $\dot{\bar{n}}=300\ {\rm quanta/s}$. 
The internal structure forms the spin system, with the states $\ket{0_s}\equiv \ket{5S_{1/2},\,m_j = -\frac{1}{2}}$, and $\ket{1_s}\equiv\ket{4D_{5/2},\,m_j = -\frac{3}{2}}$. The electronic state $\ket{2_s} \equiv \ket{5S_{1/2},\,m_j = \frac{1}{2}}$ is used as an auxiliary state.
 
The $\ket{0_s}\leftrightarrow\ket{1_s}$ quadrupole transition is driven by a 674\,nm laser; the two noncommuting spin-dependent forces required to generate the nonlinear interaction are created by two 674\,nm beams. The $\ket{2_s}$ state is isolated from the interaction, and an RF antenna drives the $\ket{0_s}\leftrightarrow\ket{2_s}$ transition.

The projective measurement is achieved by driving the \(5S_{1/2} \leftrightarrow 5P_{1/2}\) transition 
which distinguishes bright states ($\ket{0_s}$ or $\ket{2_s}$), which scatter photons, from the dark state ($\ket{1_s}$), which does not. This measurement is only non-destructive to the motional state in the case where $\ket{1_s}$ is detected, and hence no photons are actually scattered. Thus, we design our pulse sequences such that the desired oscillator state is heralded upon $\ket{1_s}$, i.e., the desired harmonic superposition is correlated with $\ket{1_s}$ just before the measurement. For example, for the state described in Eq.~\eqref{eq:qubit_entangled_superpos}, we would select the even superposition $\ket{\psi_{k,+}}$.
The projective measurement is used mid-circuit for selecting the desired superposition as well as for the final detection step. In both cases, we collect photons for \SI{200}{\micro \second}, yielding a combined dark state preparation and measurement fidelity of 0.993~\cite{supplementary}.

We create both even and odd superpositions between the state $\ket{\zeta_k}$ and $\ket{-\zeta_k}$ as illustrated in Eq.~\eqref{eq:qubit_entangled_superpos} and Fig.~\ref{fig:scheme_equal}. By measuring the qubit immediately after this creation step, we obtain the probability of collapsing in $\ket{0_s}$ and $\ket{1_s}$ respectively and, hence, the probability of generating the even and odd superposition. For example, for $k=2$ in Eq.~\eqref{eq:int_hamiltonian}, \eqref{eq:qubit_entangled_superpos} we obtain the superposition between two oscillator states squeezed about orthogonal axes. The corresponding probabilities for the even ($\ket{\psi_{2,+}}$) and odd ($\ket{\psi_{2,-}}$) states are given by ${|N_{2,\pm}|^2 = \left(2\pm 2/\sqrt{\cosh(2|\zeta_2|)}\right)/4}$, which depend on the magnitude of the squeezing parameter $|\zeta_2|$ (see Supplement ~\cite{supplementary}).
In the limit of small squeezing parameter $|\zeta_2|$, the generation of the odd superposition vanishes, which indicates that the overlap between $\ket{\zeta_k}$ and $\ket{-\zeta_k}$ tends to $1$ as the oscillator state is unchanged from the initial state.
In the limit $|\zeta_2| \rightarrow \infty$, the odd and even state have equal generation probability of $1/2$ which indicates that $\ket{\zeta_2}$ and $\ket{-\zeta_2}$ are orthogonal and their overlap is $0$. We confirm this behaviour experimentally; the data in Fig.~\ref{fig:equal_superpositions}A agree well with the theoretical prediction from an independent estimate of $|\zeta_2|$. 

Extending this protocol to higher-order nonlinear interactions, we reconstruct the Wigner quasi-probability distribution of even and odd superpositions for squeezing ($k=2$), trisqueezing ($k=3$), quadsqueezing ($k=4$) (Fig.~\ref{fig:equal_superpositions} B-G). To do so, we measure the characteristic function $\chi(\beta) = \braket{\hat{D}(\beta)}_{osc}$ of the oscillator state and infer the Wigner distribution via a Fourier transform~\cite{fluhmann2020direct, bazavan2024squeezing}. The characteristic function $\chi(\beta)$ is measured via the spin using a spin-dependent displacement $\hat{D}(\hat{\sigma}_y\beta/2)$ as shown in the tomography step in Fig.~\ref{fig:scheme_equal}A. By varying the complex-valued displacement parameter $\beta$, we can sample different values of $\chi(\beta)$ (see Supplement~\cite{supplementary}).   
 We find good agreement between the experimental results and numerical simulations using independently measured parameters (Fig.~\ref{fig:equal_superpositions} H-M). In the Supplement~\cite{supplementary}, we show the experimental data and numeric predictions of $|N_{k,\pm}|^2$ as a function of $|\zeta_k|$ for $k \in \{3,4\}$.

The superpositions created, especially the odd ones, have a large amount of Wigner negativity. In continuous variable quantum computation, large Wigner negativities are a requirement for quantum states that yield an advantage over classical computation~\cite{mari2012positive,hahn2024classical} and hence highly sought after. 
We quantify the resourcefulness of these states by evaluating their Wigner logarithmic negativity (WLN)~\cite{albarelli2018resource, takagi2018convex}, neglecting decoherence effects, and benchmark them against Fock and cat states. 
We find that for a fixed average Fock state occupation, the superpositions created in this work exhibit larger WLN than Fock or cat states~\cite{LWN_footnote} (see Supplement~\cite{supplementary}).

Another important feature is that the non-vanishing Fock state occupations of the superpositions are spaced by ${2k}$, where $k$ is the order of the interaction (see Supplement~\cite{supplementary}). As $k$ increases, so does the spacing, and the states have increased rotational symmetries as seen in Fig.~\ref{fig:equal_superpositions}. These rotational symmetries can be used to encode more robust logical qubits~\cite{grimsmo2020quantum}. For example, a single phonon loss in a (two-component) cat-qubit encoding results in an irreparable bit-flip error~\cite{deMatos1996even}. However, when the same loss channel acts on a qubit encoding formed by a superposition of squeezed states, it projects outside the code space, making the error correctable. 

Thus far, the superpositions have only involved constituents generated from the same nonlinear interactions. To create arbitrary superpositions with nonclassical constituents in the most general form 
\begin{equation}\label{eq:arbitrary}
    \ket{\psi}  \propto a \ket{\zeta_k} + b \ket{\zeta_{k'}},
\end{equation}
we require individual control of $a,b,\zeta_k, \zeta_{k'} \in \mathbb{C}$. To create this state, we extend the Hilbert space of the spin to a qutrit (three-level system), $\{\ket{0_s},\ket{1_s},\ket{2_s}\}$ (see Fig.~\ref{fig:arbitrary_superpos}A, B).
We create the constituent states via the same interaction described in \eqref{eq:int_hamiltonian}, which only couples to $\ket{0_s}$ when applied to the subspace $\{\ket{0_s}, \ket{2_s}\}$. Thus, we utilize this feature to temporarily hide one of the constituents $\ket{\zeta_k,2_s}$ while creating the other using a separate interaction~\cite{reducedtreatment}. 

The pulse sequence for creating the arbitrary superpositions is described in Fig.~\ref{fig:arbitrary_superpos}A; for simplicity, we first focus on the case $a=b$.
After the first nonlinear interaction is applied, the hybrid system is in a state $\ket{0_{osc},2_s} + \ket{\zeta_k, 0_s}$. A subsequent $R_y(\pi)_{02}$ rotation, where the indices indicate the two spin sublevels $\ket{0_s} \leftrightarrow\ket{2_s}$ coupled, then swaps the spin states, hiding the state $\ket{\zeta_k, 2_s}$ from the effect of the second nonlinear interaction. The second nonlinear interaction then creates the state $\ket{\zeta_k,2_s} + \ket{\zeta_{k'}', 0_s}$. One final $R_y(\pi/2)_{02}$ pulse followed by the shelving $R_y(\pi)_{01}$ pulse creates a state
described by Eq.~\eqref{eq:qubit_entangled_superpos}, with $\ket{\psi_{k,\pm}} \propto \ket{\zeta_k} \pm \ket{\zeta_{k'}'}$ where $\zeta_{k'}'$ is no longer $-\zeta_k$.

We then characterise the state using the same techniques as before. As mentioned previously, the unitarity of the nonlinear interaction is essential to prevent the destruction of the intermediate state by subsequent applications of the nonlinear interaction.
Splitting the sequence into two parts allows us to change the second nonlinear interaction relative to the first one~\cite{ramseyspinecho}.

In Fig.~\ref{fig:arbitrary_superpos}C, we show that we can create a superposition of squeezed states with variable orientation of the respective squeezing axis, i.e. $\zeta_{2}'=e^{i\, 2\phi}\zeta_{2}$, by adjusting the phase $\phi$ of the nonlinear interaction in the second part of the sequence relative to the first part. Hence, the created state takes the form $\ket{\zeta_2}+\ket{e^{i\, 2\phi}\zeta_2}$. We measure the value of the characteristic function at a fixed magnitude $|\beta|$ but vary its complex phase $\arg(\beta)$. At $\phi=\pi/2$ we create a superposition of states 
squeezed about orthogonal axes (equal superposition of squeezed states shown in Fig.~\ref{fig:equal_superpositions}B, H), which gives rise to four equally spaced maxima in $\chi$ when scanning $\arg(\beta)$. By scanning the degree of freedom $\phi$, the two orthogonal squeezing axes gradually merge into a single one at $\phi=\{0, \pi\}$ (regular squeezed state), giving rise to two equally spaced maxima. This behaviour results in the sawtooth pattern displayed in Fig.~\ref{fig:arbitrary_superpos}C.

In Fig.~\ref{fig:arbitrary_superpos}D, we show that we can control the magnitude of squeezing of each constituent independently~\cite{power_equal_footnote}.
We keep the squeezing parameter $\zeta_2$ of the first constituent fixed while adjusting the magnitude of the second constituent by a factor $c \in \mathbb{R}$ which results in a state $\ket{\zeta_2}+\ket{c\cdot \zeta_2}$ i.e. $\zeta_{2}'=c\zeta_{2}$. For $c=0$, this corresponds to a superposition of a squeezed state with the initial ground state, whereas for $c=-1$, this corresponds to the equal superposition shown in Fig.~\ref{fig:equal_superpositions}B, H.
We measure the variance of the state along the squeezing axis of the first constituent and an axis orthogonal to it (i.e., two orthogonal quadratures). We compute the variances using the reconstructed Wigner functions for different settings of $c$. As shown in Fig.~\ref{fig:arbitrary_superpos}D, the variance closely follows the simulated prediction.

The variance of each constituent does not vary independently. In fact, altering the magnitude of the second constituent affects the variance of the resulting state in both quadratures: the quadrature dominated by the second constituent (blue) and the quadrature dominated by the first constituent (green), even though the first constituent's squeezing magnitude remains fixed. This behaviour is a direct consequence of the interference between both constituents.

To demonstrate that we can change the order $k$ of the nonlinear interaction for each constituent, we combine a squeezed state ($k=2$) as the first constituent with a trisqueezed state ($k=3$) as the second constituent. Fig.~\ref{fig:arbitrary_superpos}E shows the resulting Wigner function of this superposition state.
\begin{figure}[!ht]
\includegraphics[]{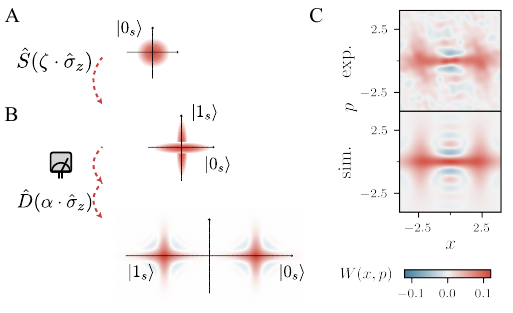}
\caption{Constructing arbitrary cat states with multiple mid-circuit measurements. ({\bf A}-{\bf B}) We illustrate the main steps under which the initial state $\ket{0_{osc}, 0_s}$ is transformed. For simplicity, we omit the single-qubit rotations before each detection. ({\bf A}) The first step is the same as in Fig.~\ref{fig:equal_superpositions}A. The hybrid system is brought into an equal superposition of squeezed states entangled with the spin.
({\bf B}) After a successful detection step we obtain the motional state $\ket{\psi_{2,+}} = \ket{\zeta_2} +\ket{-\zeta_2}$ that is no longer entangled with the spin and we can apply a second spin-conditioned operation. As an example, we choose a spin-dependent displacement, which, after a second mid-circuit step, creates a state reminiscent of a cat state where each of the cat constituents is formed by a squeezed superposition displaced in opposite directions. ({\bf C}) Experimental and simulated Wigner function of this $\psi_+$-cat state. The estimated squeezing magnitude is $|\zeta_2| = 1.25(4)$ followed by the displacement magnitude $|\alpha| = |\zeta_1|/2= 1.62(4)$. The displacement axis is aligned with one of the squeezing axis.
The experimental parameters are tabulated in the Supplement~\cite{supplementary}.}
\label{fig:concatenate_sequence}
\end{figure}
While we do not include the data in the main text, we discuss in the Supplement~\cite{supplementary} that the sequence can be amended to vary the relative amplitude and phase of the probability coefficients $a$ and $b$. By replacing the first $R_y(\pi/2)$ rotation by $R_y(\theta)$ to vary the probability amplitude ratio between the two constituents $\cos(\theta/2)\ket{\zeta_k} + \sin(\theta/2)\ket{-\zeta_k}$. Further, it is also possible to replace the $R_y(\pi)$ rotation with a rotation about an axis $\hat{\sigma}_\gamma = \cos(\gamma)\hat{\sigma}_x + \sin(\gamma)\hat{\sigma}_y$ to introduce a complex phase factor between the probability amplitude of the two constituents $\ket{\zeta_k} + \ket{-\zeta_k} e^{i2\gamma}$. Thus, we demonstrate complete control over $a, b, \zeta_k$ and $\zeta_{k'}'$ in Eq.~\eqref{eq:arbitrary}. 

Finally, we show that the sequence depicted in Fig.~\ref{fig:scheme_equal}A can be extended further to construct spatially separated superposition states. Aside from a foundational quantum optics interest, such spatially-separated superpositions are essential in various applications, including continuous variable quantum error correction codes (e.g., cat~\cite{deMatos1996even} or GKP~\cite{gottesman2001encoding} states), and metrology~\cite{gilmore2021quantum}. We create these states using an additional spin-dependent displacement followed by a mid-circuit measurement as shown in Fig.~\ref{fig:concatenate_sequence}. The second mid-circuit detection disentangles the spin from the harmonic oscillator, projecting the latter into a cat-like state, where the two spatially displaced states are themselves nonclassical and non-Gaussian. We use this protocol to create a spatially separated superposition of squeezed states 
\begin{equation}\label{eq:cat}
    \ket{\psi} \propto D(\alpha)\ket{\psi_{2,+}}+D(-\alpha)\ket{\psi_{2,+}},
\end{equation}
where $\ket{\psi_{2,+}} \propto \ket{\zeta_2}+ \ket{-\zeta_2}$.
Thus, we show that we can create an even broader class of oscillator superpositions by concatenating additional spin-dependent interactions and mid-circuit measurements.

Our work demonstrates the versatility of hybrid oscillator-spin systems in generating superpositions of oscillator states. By integrating nonlinear spin-dependent interactions with mid-circuit measurements, we have established arbitrary control of the phase, amplitude, and interaction type for the constituents of the superpositions. Using these techniques, we have experimentally demonstrated many nonclassical states that have until now only been explored theoretically~\cite{sanders1989superposition, barbosa2000generalised, nicacio2010phase,drechsler2020state,ayyash2024driven}. Expanding complex harmonic oscillator states from cat~\cite{deMatos1996even} and GKP~\cite{gottesman2001encoding} states to superpositions of squeezed or superpositions of non-Gaussian states is not only of foundational interest but opens new avenues for error correction, sensing protocols, and continuous variable or hybrid oscillator-spin quantum computing~\cite{liu2024hybrid}. For example, these superposition states have increased rotational symmetries, which enable robust error correction~\cite{grimsmo2020quantum}.  Further, their large Wigner negativities indicate that they generally cannot be efficiently simulated classically~\cite{mari2012positive, hahn2024classical}. Their utility applies not only to quantum computation but also to quantum-enhanced metrology. For example, the squeezed superposition state is suitable as a displacement sensor~\cite{drechsler2020state}, which, in the case of trapped ions, can be used for sensing small electric fields~\cite{gilmore2021quantum}.
The demonstrated techniques can be extended to multiple oscillators by coupling to additional motional modes of the ion, for example, to create two-mode squeezed superpositions~\cite{superposition2021cardoso} or in general to increase the size of the hybrid quantum system. Aside from trapped ions, the tools developed here can be applied to any physical system with a quantum harmonic oscillator coupled to a spin such as superconducting circuits~\cite{wallraff2004strong}, nanoparticles~\cite{delord2020spin}, or atoms coupled to a cavity~\cite{hacker2019deterministic} or in optical tweezers~\cite{kaufman2012cooling}. For systems with more massive oscillators, these superpositions could be used to test not only the boundaries of the classical and quantum world~\cite{nimmrichter2013macroscopicity}, but how quantum physics interacts with gravity~\cite{penrose1996on}. 


\section*{Acknowledgements}
We would like to thank Alejandro Bermudez, Alexander Lvovsky, Andrew Daley, Joshua Combes and his research group, Mattia Walschaers, and Scott Parkins for very insightful discussions and comments on the manuscript.
This work was supported by the US Army Research Office (W911NF-20-1-0038) and the UK EPSRC Hub in Quantum Computing and Simulation (EP/T001062/1). GA acknowledges support from Wolfson College, Oxford. CJB acknowledges support from a UKRI FL Fellowship. RS acknowledges funding from the EPSRC Fellowship EP/W028026/1 and Balliol College, Oxford. 

\section*{Author contributions}
SS and OB led the experiments and analysed the results
with assistance from DJW, GA, and RS; SS, OB, DJW, and GA updated and maintained the experimental apparatus; SS, OB, and RS conceived the experiments and wrote the manuscript with input from all authors; RS supervised the work with support from DML and CJB; DML and CJB secured funding for the work.

\section*{Competing Interests}
RS is partially employed by Oxford Ionics Ltd. CJB is a director of Oxford Ionics Ltd. All other authors declare no competing interests.

\section*{Data availability}
Source data for all plots and analysis code that support the plots are available from the corresponding authors upon reasonable request.

\bibliographystyle{apsrev4-2}
\bibliography{bibliography}

\begin{thebibliography}{62}
\bibitem [{LWN()}]{LWN_footnote}%
  \BibitemOpen
  \href@noop {} {}\bibinfo {note} {While LWN is a resource monotone, it is still an active area of research under which circumstances it adequately quantifies the difficulty of simulating the state classically.}\BibitemShut {Stop}%
\bibitem [{red()}]{reducedtreatment}%
  \BibitemOpen
  \href@noop {} {}\bibinfo {note} {The treatment can be reduced to the subspace $\{\ket{0_s}, \ket{2_s}\}$ because the interaction commutes with the computational basis and hence does not cross-couple the levels.}\BibitemShut {Stop}%
\bibitem [{ram()}]{ramseyspinecho}%
  \BibitemOpen
  \href@noop {} {}\bibinfo {note} {This sequence can be understood as a Ramsey-spin echo sequence on the subsystem $\{\ket{0_s}, \ket{2_s}\}$ and indeed protects us against uncontrolled frequency shifts between the sublevels too.}\BibitemShut {Stop}%
\end{thebibliography}


\clearpage
\renewcommand{\thefigure}{B.\arabic{figure}}
\setcounter{figure}{0}
\renewcommand{\theequation}{A.\arabic{equation}}
\setcounter{equation}{0}
\setcounter{section}{0}
\onecolumngrid
\begin{center}
\vspace{5 mm}
\textbf{\large Supplemental Material for:\\
Generating arbitrary superpositions of nonclassical quantum harmonic oscillator states}\\
\end{center}
\twocolumngrid

\section{Theory}
\subsection{Generating the nonlinear spin-dependent interaction}
While the method employed to generate the nonlinear interaction is described in detail in Refs.~\citenum{sutherland2021universal,bazavan2024squeezing}. We summarise the relevant details below.
We combine two spin-dependent forces (SDF) with non-commuting spin-conditionings $[\hat{\sigma}_\alpha, \hat{\sigma}_{\alpha'}] \neq 0$ and $\phi$ the oscillator phase of the second SDF which results in a Hamiltonian
\begin{equation}
    \begin{split}
    \hat{H} &= \frac{\hbar\Omega_\alpha}{2} \hat{\sigma}_\alpha (\hat{a}\, e^{-i\Delta t}+ \hat{a}^\dagger e^{i\Delta t})\\
    &+\frac{\hbar\Omega_{\alpha'}}{2} \hat{\sigma}_{\alpha'} (\hat{a}\, e^{-i (m\Delta\, t+\phi)} + \hat{a}^\dagger e^{i(m\Delta\, t+ \phi)}).
    \end{split}
\end{equation}
The two forces are detuned from resonance with the oscillator by $\Delta$ and $m\Delta$, respectively. By choosing $k = 1-m$, we can resonantly drive effective generalised squeezing interactions
\begin{equation}
    \hat{H}_k =\frac{{\hbar}\Omega_k}{2} \hat{\sigma}_\beta (\hat{a}^k e^{-i\phi} + (\hat{a}^\dagger)^k e^{i\phi}),
\end{equation}
which for $\hat{\sigma}_\beta = \hat{\sigma}_z$ corresponds to Eq.~\eqref{eq:int_hamiltonian} in the main text, where
\begin{align}
    \hat{\sigma}_\beta &\propto 
    \begin{cases}
    [\hat{\sigma}_{\alpha}, \hat{\sigma}_{\alpha'}]& \text{if } k~{\rm mod}~2=0\\
    \hat{\sigma}_{\alpha'}              & \text{otherwise.}
\end{cases}
\end{align}
While we can choose our spin bases such that $\hat{\sigma}_\beta = \hat{\sigma}_z$, and do so for the majority of the data,
it was convenient for Fig.~\ref{fig:equal_superpositions}DE to choose $\hat{\sigma}_\beta = \hat{\sigma}_x$ which maximises the strength and avoids two single qubit rotations. We have indicated the native spin-conditioning for every sequence in Sec.~\ref{supp:sec_exp_params}. 
The coupling strength $\Omega_k$ 
\begin{equation}
    \Omega_{2,3,4} = \Biggl\{  \frac{\Omega_{\alpha'} \Omega_{\alpha}}{\Delta}, \frac{\Omega_{\alpha'} \Omega_{\alpha}^2}{2 \Delta^2},\frac{\Omega_{\alpha'} \Omega_{\alpha}^3}{8 \Delta^3} \Biggl\} \cdot \sin(\theta_{\alpha,\alpha'}),
\end{equation}
which is proportional to $\sin(\theta_{\alpha, \alpha'})$ where $\theta_{\alpha,\alpha'}$ denotes the angle between the spin bases $\alpha$ and $\alpha'$ and is maximised if the two spin bases are orthogonal. We utilise this angle to adjust the squeezing magnitude for the constituent $\ket{c\cdot \zeta_2}$ in the superposition of Fig.~\ref{fig:arbitrary_superpos}D, while keeping the optical power and the duration of the interaction fixed.
For the quantum circuits presented in the main text, we adjust $\Delta$, $\Omega_\alpha$, $\Omega_{\alpha'}$, and $\hat{\sigma}_\beta$. The settings are tabulated in Sec.~\ref{supp:sec_exp_params}.  

\subsection{Normalisation coefficient}
We require the squeezed state superposition to be normalised i.e. $\braket{\psi_\pm |\psi_\pm} = 1$, and find a normalisation coefficient
\begin{align}
    \begin{split}
    |N_{2,\pm}|^2 &= \frac{1}{4}(\bra{\zeta_2} \pm \bra{\zeta_2})(\ket{\zeta_2}\pm \ket{\zeta_2})\\
    &=\frac{1}{4} \braket{\zeta_2|\zeta_2}+  \braket{-\zeta_2|-\zeta_2} \pm 2{\rm Re}(\braket{-\zeta_2|\zeta_2})\\
    &= \left(2 \pm \frac{2}{\sqrt{\cosh(2|\zeta_2|)}}\right)/4,
    \end{split}
\end{align}
where we assume that each squeezed state is normalised. The inner product $\braket{-\zeta_2|\zeta_2}$ can be determined from the Fock state expression of a vacuum squeezed state~\cite{stoler1970equivalence, yuen1976two,weedbrook2012gaussian}:
\begin{align}
    \ket{\zeta_2} &= \frac{1}{\sqrt{\cosh(|\zeta_2|)}} \sum_{n=0}^\infty (- e^{i \arg(\zeta_2)} \tanh(|\zeta_2|))^n \frac{\sqrt{(2n)!}}{2^n n!} \ket{2n},\\
    \begin{split}
    &\braket{-\zeta_2|\zeta_2} = \frac{1}{\cosh(|\zeta_2|)} \sum_{n=0}^\infty e^{i \pi n}\tanh(|\zeta_2|)^{2n} \frac{(2n)!}{2^{2n} (n!)^2} \\
    &\phantom{\braket{-\zeta_2|\zeta_2}}= \frac{1}{\cosh(|\zeta_2|)} \frac{1}{\sqrt{\tanh(|\zeta_2|)^2 +1 } } = \frac{1}{\sqrt{\cosh(2|\zeta_2|)}}. 
    \end{split}
\end{align}

For the superpositions involving non-Gaussian (i.e.~trisqueezed or quadsqueezed) constituents, it is unclear if closed-form solutions exist~\cite{braunstein1987generalised}.

\subsection{Native gate sequences}

In this section, we give a more detailed description of the pulse sequence used to generate the superposition states. For simplicity, we omitted the spin-echo $R(\pi)$ pulses described below from the main text and Fig.~\ref{fig:scheme_equal}.

All nonlinear interactions $\hat{U}_{\rm NL}$ except for trisqueezing in Fig.~\ref{fig:equal_superpositions} are conditioned on $\hat{\sigma}_z= \ket{1_s}\bra{1_s} - \ket{0_s}\bra{0_s}$.
For squeezing and quadsqueezing, we set $\hat{\sigma}_{\alpha} = \sigma_{x}$ and $\sigma_{\alpha'}= \hat{\sigma}_y$ which results in $\hat{\sigma}_\beta = \hat{\sigma}_z$. 
For the trisqueezing superposition shown in Fig.~\ref{fig:equal_superpositions}, we keep this setting which results in the spin basis of the nonlinear interaction $\hat{\sigma}_\beta = \hat{\sigma}_x$.
For trisqueezing shown in Fig.~\ref{fig:arbitrary_superpos}, we change the trisqueezing spin basis compared to Fig.~\ref{fig:equal_superpositions}. We set the spin conditioning of the SDF to $\hat{\sigma}_{\alpha'}=\hat{\sigma}_z$ such that the resulting nonlinear interaction is conditioned on $\hat{\sigma}_\beta=\hat{\sigma}_z$.

\begin{align}
    \ket{0_{osc},0_s} &\stackrel{R_y(\pi/2)}{\rightarrow}\frac{\ket{0_{osc}, 0_s} +\ket{0_{osc}, 1_s}}{\sqrt{2}} \stackrel{\hat{U}_{\rm NL}}{\rightarrow} \frac{\ket{\zeta/2, 0_s} + \ket{-\zeta/2, 1_s}}{\sqrt{2}} \label{eq:supp_first_arm_ramsey}\\
    &\stackrel{R_y(\pi)}{\rightarrow} \frac{-\ket{\zeta/2, 1_s} + \ket{-\zeta/2, 0_s}}{\sqrt{2}}\stackrel{\hat{U}_{\rm NL}|_\pi}{\rightarrow} \frac{-\ket{\zeta, 1_s} + \ket{-\zeta, 0_s}}{\sqrt{2}}\\
    &\stackrel{R_y(\pi/2)}{\rightarrow} - N_+ (\ket{\zeta} +\ket{-\zeta}) \ket{0_s} + N_- (\ket{\zeta} -\ket{-\zeta}) \ket{1_s}.
\end{align}
We can adjust the phases of the qubit rotations to adjust which harmonic oscillator superposition is heralded by $\ket{1_s}$. For example, we can change the axis of the $R_y(\pi)$ pulse to $R_x(\pi)$. The sequence continued from Eq.~\eqref{eq:supp_first_arm_ramsey} then looks like
\begin{align}
&\stackrel{R_x(\pi)}{\rightarrow} \frac{-i\ket{\zeta/2, 1_s} -i\ket{-\zeta/2, 0_s}}{\sqrt{2}}\stackrel{\hat{U}_{\rm NL}|_\pi}{\rightarrow} \frac{-i\ket{\zeta, 1_s} - i\ket{-\zeta, 0_s}}{\sqrt{2}}\\
    &\stackrel{R_y(\pi/2)}{\rightarrow} i N_- (\ket{\zeta} -\ket{-\zeta}) \ket{0_s} - i N_+ (\ket{\zeta} +\ket{-\zeta}) \ket{1_s}.
\end{align}

For the trisqueezing superposition shown in Fig.~\ref{fig:equal_superpositions}, we omit the $R(\pi/2)$ and $R(\pi)$ pulses.
To herald the even (odd) superposition upon detecting $\ket{1_s}$, we initialise the starting state in $\ket{1_s}$ ($\ket{0_s}$)  before applying the interaction.

For the sequence in Fig.~\ref{fig:arbitrary_superpos}, we require all of the applied nonlinear interaction to be conditioned on $\hat{\sigma}_z$ such that we can treat its impact on the subsystem $\{\ket{0_s}, \ket{2_s}\}$ as a closed system. 

\subsection{Adjusting the probability amplitude coefficient of the superpositions}

\begin{figure}[!ht]
\includegraphics[width=\linewidth]{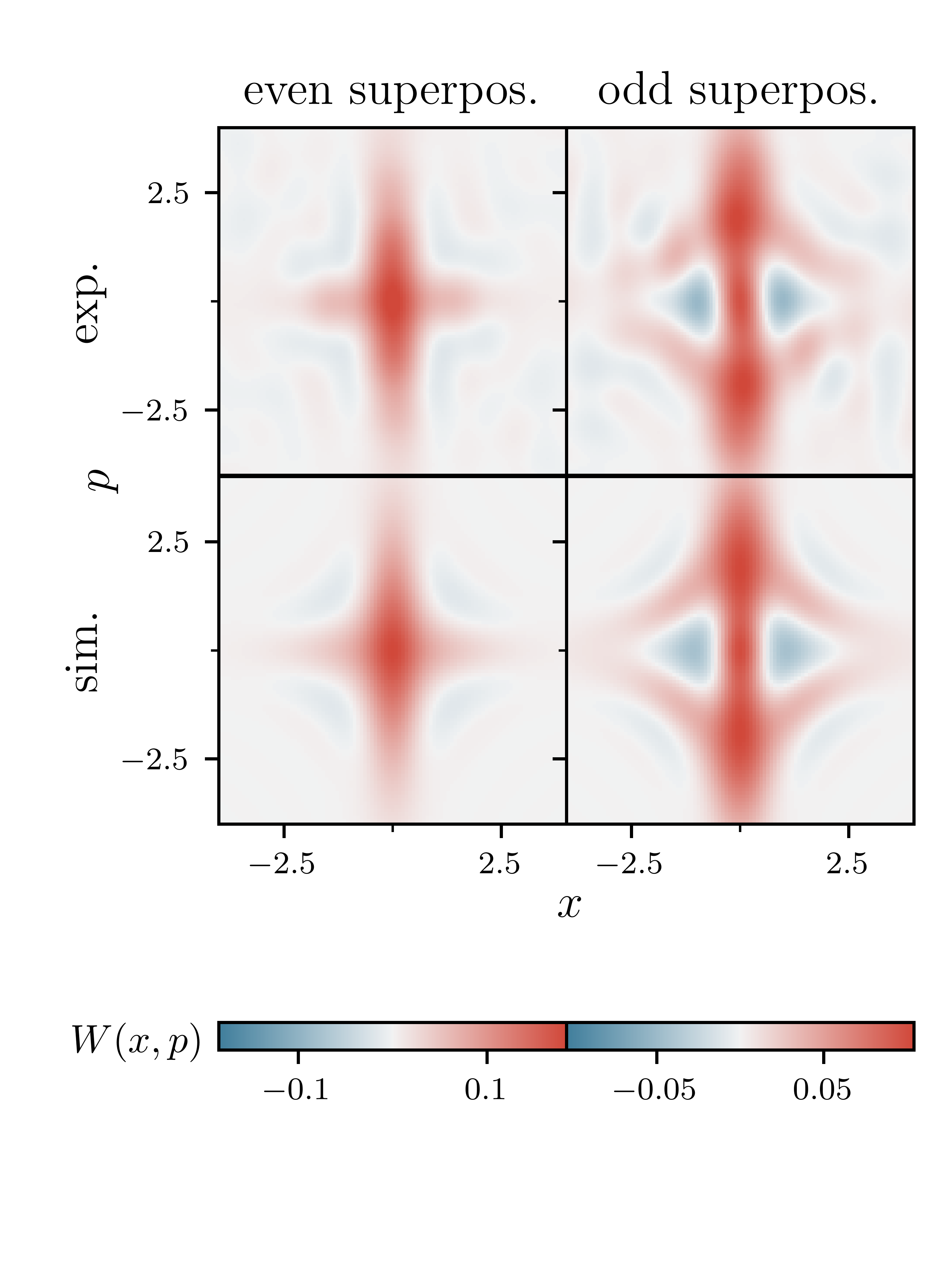}
\caption{Wigner function of the state $\ket{\psi} \propto a \ket{\zeta_2} \pm b\ket{-\zeta_2}$ with a probability amplitude ratio $a/b = \cos(\theta/2)/\sin(\theta/2) = \sqrt{2}+1$ where $\theta= \pi/4$. We show experimental and numerical simulation data for both even (+) and odd (-) superpositions. We choose $|\zeta_2|=1.12(5)$, consistent with Fig.~\ref{fig:equal_superpositions}B.}
\label{fig:supp_complex_amplitude_magnitude}
\end{figure}

We can arbitrarily set the probability amplitudes of each constituent instead of only balanced superpositions. 
To adjust the relative amplitudes, we can replace the first $R_y(\pi/2)$ rotation by $R_y(\theta)$ rotation such that the initial spin superposition is the form 
\begin{equation}
    \ket{\psi} = \cos(\theta/2)\ket{0_{osc}, 0_s} + \sin(\theta/2)\ket{0_{osc}, 1_s}.
\end{equation}
The rest of the sequence remains unchanged. This results in an amplitude ratio $\cos(\theta/2)/\sin(\theta/2)$ between the constituents of the superposition. For example, for $\theta=\pi/2$, the amplitudes are equal and the sequence is the same as in Eq.~\ref{eq:supp_first_arm_ramsey}. 
As a proof of principle, we produce a squeezed state superposition with $\theta=\pi/4$ i.e.~$a \ket{\zeta_2} + b\ket{-\zeta_2}$ with an amplitude ratio $a/b = \sqrt{2}+1$ (see Fig.~\ref{fig:supp_complex_amplitude_magnitude}).

To adjust the relative phase of the complex probability amplitudes, we vary the axis $\hat{\sigma}_\gamma = \cos(\gamma)\hat{\sigma}_x + \sin(\gamma)\hat{\sigma}_y$ of the $R_\gamma(\pi)$ rotation. After the full sequence, we obtain
\begin{align}
    \begin{split}
    \ket{\psi} &= -N_{-}^{(\gamma)}\left(\ket{\zeta}i e^{i\gamma} -\ket{-\zeta}ie^{-i\gamma}\right)\ket{0_s} \\
    &+N_{+}^{(\gamma)} \left(\ket{\zeta}i e^{i\gamma} +\ket{-\zeta}ie^{-i\gamma}\right)\ket{1_s}.
    \end{split}
\end{align}

 After rearranging, we obtain
\begin{align}
    \begin{split}
    \ket{\psi} = -ie^{i\gamma} N_{-}^{(\gamma)}\left(\ket{\zeta} -\ket{-\zeta}e^{-i2\gamma}\right)\ket{0_s} \\
    + ie^{i\gamma} N_{+}^{(\gamma)} \left(\ket{\zeta} +\ket{-\zeta}e^{-i2\gamma}\right)\ket{1_s}.
    \end{split}
\end{align}

\begin{figure}[!ht]
\includegraphics[width=\linewidth]{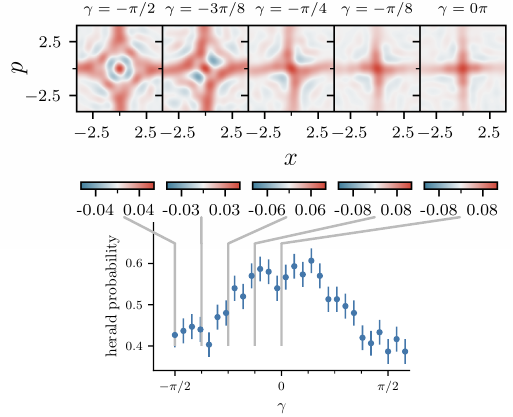}
\caption{Measuring the herald probability $|N_+^{(\gamma)}|^2$ and the resulting Wigner functions of the state $\ket{\psi}_+^{(\gamma)} \propto \ket{\zeta_2} + \ket{-\zeta_2}e^{i2\gamma}$ as a function of $\gamma$. The Wigner function gradually transforms from the even into the odd superposition. The detection probabilities change accordingly. We choose $|\zeta_2|=1.67(7)$ consistent with Fig.~\ref{fig:equal_superpositions}C. The error bars for $|N_+^{(\gamma)}|^2$ denote the 68\% confidence interval.}
\label{fig:supp_complex_amplitude_phase}
\end{figure}

In Fig.~\ref{fig:supp_complex_amplitude_phase}, we show the herald probability $|N_{+}^{(\gamma)}|^2$ and the resulting Wigner functions when changing the relative phase of the probability amplitude $\gamma$.

\subsection{Measuring the variance of squeezed superpositions}
To estimate the variance in Fig.~\ref{fig:arbitrary_superpos}D based on the experimentally reconstructed Wigner function $W$, we estimate the probability distribution $P_x$ and $P_p$ which can be found by integrating the Wigner function along the orthogonal quadratures $p$ and $x$ respectively:
\begin{align}
    P_x(x) &= \int  W(x,p) dp,\\
    P_p(p) &= \int  W(x,p) dx.
\end{align}
The variances are then
\begin{align}
    {\rm Var}(\hat{x}) &= \braket{\hat{x}^2} - \braket{\hat{x}}^2,\\
    {\rm Var}(\hat{p}) &= \braket{\hat{p}^2} - \braket{\hat{p}}^2,
\end{align}
where the moments of $x$ and $p$ are given by
\begin{align}
    \braket{\hat{x}^k} &= \int x^k P_x(x) dx,\\
    \braket{\hat{p}^k} &= \int p^k P_p(p) dp.
\end{align}
For the experimental data, we rotate the matrix representing the discretised Wigner function such that the principle axes are aligned with $x$ and $p$. For the simulated data, we extract a discretised Wigner function before performing the same analysis. We calculate the probability distributions and the moments by approximating the integration via the trapezoidal rule. 
\section{Extended experimental data and parameters}
\subsection{Experimental parameters for the results in the main text}\label{supp:sec_exp_params}
For Figs.~\ref{fig:equal_superpositions}, \ref{fig:arbitrary_superpos}, and \ref{fig:concatenate_sequence} we used the experimental parameters tabulated in Tab.~\ref{tab:supp_equal_superpos_params}.
\begin{table*}[ht]
\centering
\caption{Experimental parameters for Figs.~\ref{fig:equal_superpositions},~\ref{fig:arbitrary_superpos}, and \ref{fig:concatenate_sequence}. All $\hat{\sigma}_z$ interactions are embedded in a spin echo sequence; hence the duration $t$ is given as $2\ {\rm x}\ t_{\rm arm}$ where $t_{\rm arm}$ is the full-width half maximum duration for which the interaction is applied in each arm. We show the used power $P$ at the ion per SDF, detuning $\Delta$ of the SDFs used to generate the interaction, the spin-conditioning $\hat{\sigma}_\beta$, the approximate herald probability $p= N_{\rm surv}/N_{\rm shots}$, where the number of initial shots $N_{\rm shots}$ and $N_{surv}$ the number of heralds, the estimated generalised squeezing parameter $|\zeta_k|$ and $k$ the order of the nonlinearity. For Fig.~\ref{fig:concatenate_sequence}, $|\zeta_1|= \Omega_\alpha t$.}
    \label{tab:supp_equal_superpos_params}
    \begin{ruledtabular}
    \begin{tabular}{lcccccccccc} 
        Fig. &{P (mW)} & $\Omega_{\alpha},\, \Omega_{\alpha'}$ ($2\pi\, \unit{\kilo\hertz}$)& $\Delta/2\pi$ (kHz) & $t$ (\unit{\micro \second}) & $\hat{\sigma}_\beta$  & $p = N_{\rm surv}/N_{\rm shots}$&$|\zeta_k|$&$k$ \\ \midrule
        2A & 0.5 & 0.601(2), 0.608(2) & 50 & 2 x (0 - 440) & $\hat{\sigma}_z$& -/300 & 0-2.5(1)& 2\\
        2BH & 0.5 & 0.601(2), 0.608(2) & 50 & 2 x 200& $\hat{\sigma}_z$& $\sim 340/500$ & 1.12(5) &2 \\ 
        2CI& 0.5 & 0.601(2), 0.608(2)& 50 & 2 x 300 & $\hat{\sigma}_z$ & $\sim 390/1000$ & 1.67(7)&2\\ 
        2DJ& 1& 0.849(3), 0.860(3) & 25 & 500 & $\hat{\sigma}_x$& $\sim 330/400$& 0.74(5)&3\\ 
        2EK& 1 & 0.849(3), 0.860(3) & 25 & 500 & $\hat{\sigma}_x$& $\sim 270/1000$& 0.74(5)&3\\ 
        2FL& 1 & 0.849(3), 0.860(3) & 25 & 2 x 300& $\hat{\sigma}_z$& $\sim 450/500$& 0.059(5)&4\\ 
        2GM& 2&1.201(4), 1.216(4)& 25 &2 x 200 & $\hat{\sigma}_z$& $\sim 160/1200$& 0.16(1)&4\\ 
        3CD &0.5 &0.601(2), 0.608(2)& 25 & 200/200& $\hat{\sigma}_z$& $\sim 340/500$ & 1.12(5)&2  \\ 
        3E & 2 &  1.201(4), 1.216(4)/0.363(1)& 100/-25 & 200/200 & $\hat{\sigma}_z$ & $\sim 300/400$ & 1.12(5)/0.25(2) &2/3\\
        4 &0.5 &0.608(2), 0.672(2)/-& 25/0 & 2 x 100/108.6& $\hat{\sigma}_z/ \hat{\sigma}_x$& $\sim 420/1200$ & 1.25(4)/3.23(7)& 2/1  \\
    \end{tabular}
    \end{ruledtabular}
\end{table*}

For Fig.~\ref{fig:supp_complex_amplitude_magnitude} we use the same parameter configuration as for Fig.~\ref{fig:equal_superpositions}B and adjust the number of shots such that we have more than $300$ successful heralds per point.
For Fig.~\ref{fig:supp_complex_amplitude_phase} we use the same parameter configuration as for Fig.~\ref{fig:equal_superpositions}C and adjust the number of shots such that we have more than $300$ successful heralds per point. 

\subsection{Mid-circuit detection error}
\begin{figure}[!ht]
\includegraphics[width=\linewidth]{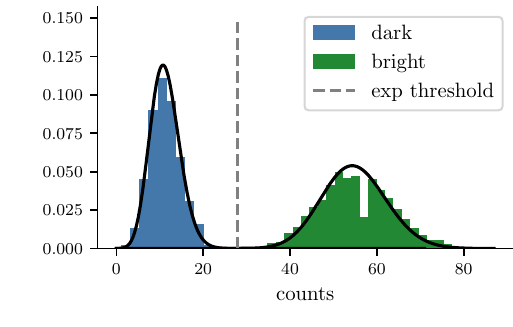}
\caption{Histogram of the fluorescence counts during a readout sequence (\SI{200}{\micro\second}) when preparing the spin in $\ket{1_s}$ (dark) or $\ket{0_s}$ (bright). The black line denotes fits to a Poisson distribution with an expected count of $11.24$ and $54.82$ 
, respectively. The experimental threshold was set to 28 counts.}
\label{fig:supp_histograms}
\end{figure}
\begin{figure}[!ht]
\includegraphics[width=\linewidth]{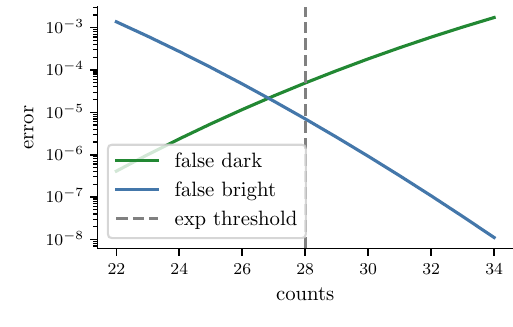}
\caption{Cumulative error probability of the Poisson fits in Fig.~\ref{fig:supp_histograms}. The left tail of the bright distribution leads to false dark counts $P({\rm dark}|{\rm bright})$. The right tail of the dark distribution leads to false bright counts $P({\rm bright}|{\rm dark}) = 1- P({\rm dark}|{\rm dark})$. With the experimental threshold of 28 counts, we have false dark and false bright probabilities of \num{4.9e-5} and \num{7.e-6}, respectively. In practice, the false bright probability is dominated by the finite lifetime of the dark state (\SI{\sim0.4}{\second}) which at a readout duration of \SI{200}{\micro\second} leads to an error of order $\sim \frac{\SI{200}{\micro\second}}{\SI{0.4}{\second}} = \SI{5e-4}{}$.   
}
\label{fig:supp_false detection}
\end{figure}
The fidelity of the mid-circuit detection directly determines the degree of certainty to which we know the heralded state. We consider the imperfect detection in the case of heralding the odd superposition $\ket{\psi_{k,-}}$. Just before detection, the wave function is 
\begin{equation}
    \ket{\psi} = N_{k,-}\ket{\psi_{k,-}}\ket{1_s} + N_{k,+}\ket{\psi_{k,+}}\ket{0_s}.
\end{equation}
With perfect detection, we could obtain the true description of $\ket{\psi_{k,-}}$. With detection errors, if we herald on $\ket{1_s}$ (dark) the state reads
\begin{equation}
\begin{split}
    \rho_{\rm detect} &\propto P({\rm dark}|{\rm dark}) |N_{k, -}|^2 \ket{\psi_{k,-}}\bra{\psi_{k,-}} \\
    &+ P({\rm dark}|{\rm bright}) |N_{k, +}|^2 \ket{\psi_{k,+}}\bra{\psi_{k,+}},
\end{split}
\end{equation}
where $P({\rm dark}|{\rm dark})$ is the detection probability of dark given dark, which is the complementary event of $P({\rm bright}|{\rm dark})$ (false bright) and $P({\rm dark}|{\rm bright})$ the probability of dark given bright (false dark).
Hence, the ratio of the mixture $m$ of making the desired state, to the false dark state is given by
\begin{equation}
    m = \frac{P({\rm dark}|{\rm dark}) |N_{k, -}|^2}{P({\rm dark}|{\rm bright})|N_{k, +}|^2 } \approx \frac{|N_{k, -}|^2}{P({\rm dark}|{\rm bright})|N_{k, +}|^2 }. 
\end{equation}
In Fig.~\ref{fig:supp_histograms}, we measure and fit the histograms of our detection sequence. From the fits (see Fig.~\ref{fig:supp_false detection}), we estimate the false dark probability to be $P({\rm dark}|{\rm bright}) = \num{4.9e-5}$. 
For the interactions considered the true herald ratio $|N_{k,-}|^2/|N_{k,+}|^2 \geq 0.1$. Hence, given our readout errors, the resulting $m > 10^3$. If the true herald ratio is even lower, the experimental threshold can be adjusted at the expense of a small loss in overall detection events.
We can also measure a cumulative state-preparation and readout (SPAM) error by calculating the dark and bright probability based on the actual histogram data. We find $P({\rm dark}) = 0.993$. Hence, the probability of preparing and measuring dark is dominated by the state preparation process.

Another error contribution in our experiment was leakage of the \SI{1033}{\nano \meter} deshelving laser. 
This results in a decay of the excited $\ket{1_s}$ state that is faster than the lifetime of the transition. This leakage is observed in the baseline measurement of Fig.~\ref{fig:equal_superpositions}A (green crosses). At \SI{880}{\micro \second}, we observe a contrast loss of $\sim 5\%$. This could be resolved by improving the extinction of the \SI{1033}{\nano \meter} when it is switched off.

\subsection{Effect of mid-circuit detection on ion motion}
\begin{figure}
    \centering
    \includegraphics{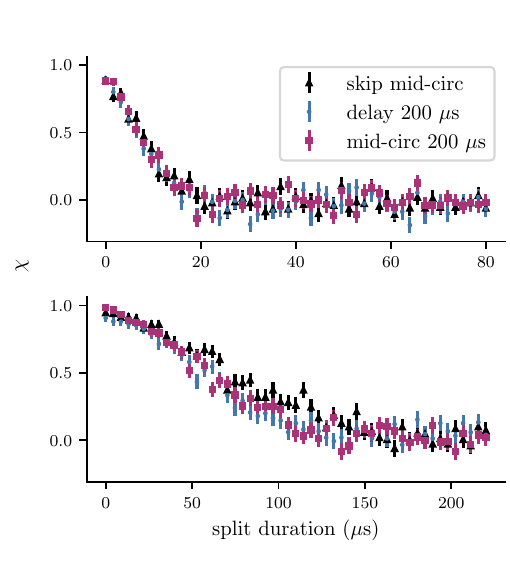}
    \caption{Real part of the characteristic function $\chi$ along the two principal axes of a squeezed state. In these experiments, we create a squeezed state before applying a spin-dependent force for a variable duration along each of the principal axes. We include three different versions of the mid-circuit operation: the full measurement, a \SI{200}{\mu\second} delay, and skipping this step altogether. Along the squeezed axis, the three settings do not differ. Along the anti-squeezed axis, the two settings that increase the sequence length by \SI{200}{\mu\second} decay a bit faster. This faster decay is a result of the non-negligible heating effects and is discussed in Ref.~\cite{bazavan2024squeezing}. The error bars denote the 68\% confidence intervals.}
    \label{fig:supp_unperturbed}
\end{figure}

We verify that performing a mid-circuit detection on a dark state ($\ket{1_s}$) leaves the motional state unperturbed. We first prepare a regular squeezed state $\ket{\zeta_2, 1_s}$, followed by the mid-circuit readout before finally measuring the characteristic function along the principle axes of the squeezed state.
We investigate three different versions of the mid-circuit measurement: regular measurement (i.e.~collect fluorescence for \SI{200}{\micro\second}), adding \SI{200}{\micro\second} delay but not applying the fluorescence laser, and omitting the mid-circuit measurement altogether. The measurement of the characteristic function for the three different settings is shown in Fig.~\ref{fig:supp_unperturbed}.  We conclude that the mid-circuit measurement does not perturb the motional state beyond increasing the sequence length by the mid-circuit measurement duration.
\subsection{Detection probabilities for equal superpositions shown in Fig.~\ref{fig:equal_superpositions}.}
\begin{figure}
    \centering
    \includegraphics{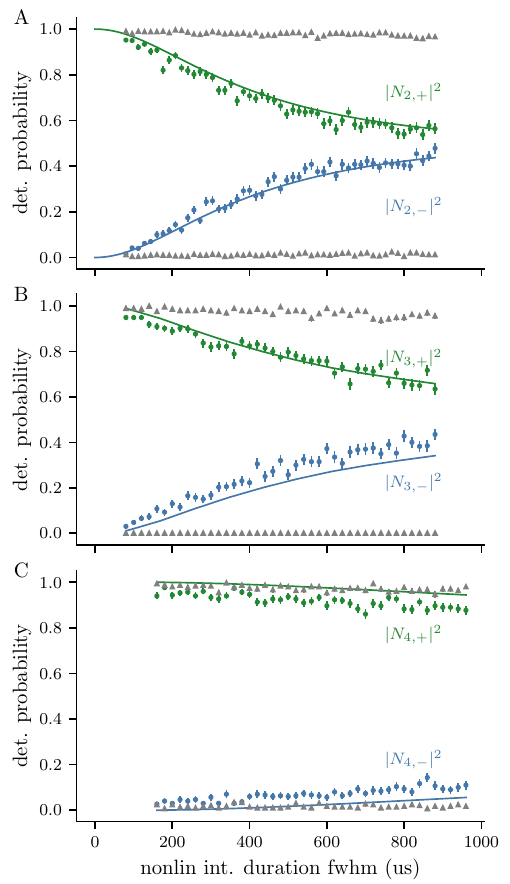}
    \caption{Probabilities of detecting the even ($|N_{k,+}|^2$, blue) and odd ($|N_{k,-}|^2$, green) superpositions $\ket{\psi_{k,\pm}} = 1/(2 N_{k,\pm})\cdot (\ket{\zeta_k}\pm \ket{-\zeta_k})$  as a function of the interaction duration. We present the data for $k=2$ (A), $k=3$ (B), and $k=4$ (C), for squeezing, trisqueezing, and quadsqueezing, respectively.  The theoretical detection probability values $|N_{2,\pm}|^2$ predicted by the coefficients in Eq.~\eqref{eq:qubit_entangled_superpos} are displayed as solid lines in ({\bf A}). 
    For ({\bf B}, {\bf C}), the solid lines represents the numerical prediction of $|N_{k,\pm}|^2$.
    Triangles denote the baseline where the system is left idle for the same duration as the non-linear interaction is applied.
    The error bars denote the 68\% confidence interval.
    Figure A is identical to Fig.~\ref{fig:equal_superpositions}A if the x-axis is rescaled to $|\zeta_2| = \Omega_2 t_{\rm fwhm}$.}
    \label{fig:supp_probability_all_int}
\end{figure}

We repeat the detection probability measurements shown in Fig.~\ref{fig:equal_superpositions}A for the trisqueezed and quadsqueezed superpositions (see Fig.~\ref{fig:supp_probability_all_int}). A closed-form analytic solution does not exist for these superpositions~\cite{braunstein1987generalised}; we replace the theory line with a numerically simulated curve. 
We compare the detection probabilities of these superpositions to the probabilities obtained when no interaction is applied (i.e.~system left idle).
Especially for the quadsqueezed state, the change in probability from leaving the system idle becomes comparable to the detection probability. This explains why the quadsqueezed superposition substantially deviates from the numerical prediction.

\section{Characterising the superpositions}
\subsection{Wigner negativity of different ideal superpositions}
There are several measures to quantify the utility, or more precisely, the resourcefulness of the oscillator state superpositions. Here, we choose the Wigner logarithmic negativity (WLN) \cite{albarelli2018resource, takagi2018convex}
\begin{equation}\label{eq:WLN}
    \mathbf{W}(\rho) = \log\left(\int dx dp|W_\rho(x,p)|\right),
\end{equation}
where $W$ is the Wigner-function of state $\rho$. 
WLN is a resource monotone, but it is still an active area of research under which circumstances it adequately quantifies the difficulty of simulating the state classically. 
We calculate the expected WLN $\mathbf{W}$ of the ideal version of the equal superposition states studied in this work (coherent, squeezed and trisqueezed superpositions) and pure Fock states using a discrete version of Eq.~$\eqref{eq:WLN}$. We measure the negativity as a function of the mean phonon number $\bar{n} = \braket{\hat{N}}_{osc}$ for the specific oscillator state.
\begin{figure}[ht]
\includegraphics[width=\linewidth]{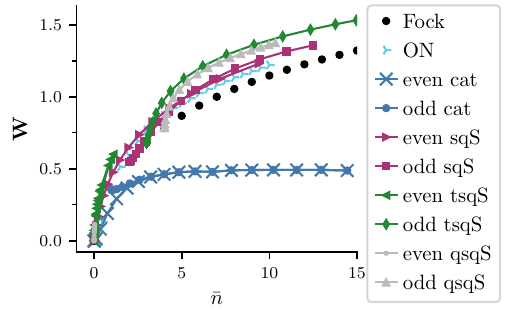}
    \caption{Wigner logarithmic negativity $\mathbf{W}$ for the states synthesised in this article as a function of mean phonon number $\bar{n} = \braket{N}_{osc}$. We calculate $\mathbf{W}$ via numerical simulation assuming a perfect noiseless system with an initial oscillator state with $\bar{n}=0$ (i.e.~perfect ground state) and no heating. We estimate $\mathbf{W}$ for Fock states, Fock state superpositions $\ket{0_{osc}}+i\ket{N_{osc}}$ (ON), cat states, squeezed superpositions (sqS), trisqueezed superpositions (tsqS) and the quadsqueezed superpositions (qsqS). The superpositions exhibit $\mathbf{W}$ exceeding the Fock and ON states for the same $\bar{n}$.}
\label{fig:supp_wigner_negativity}
\end{figure}

\subsection{Comparing the realistic and ideal states}
\begin{figure}[ht]
\includegraphics[width=\linewidth]{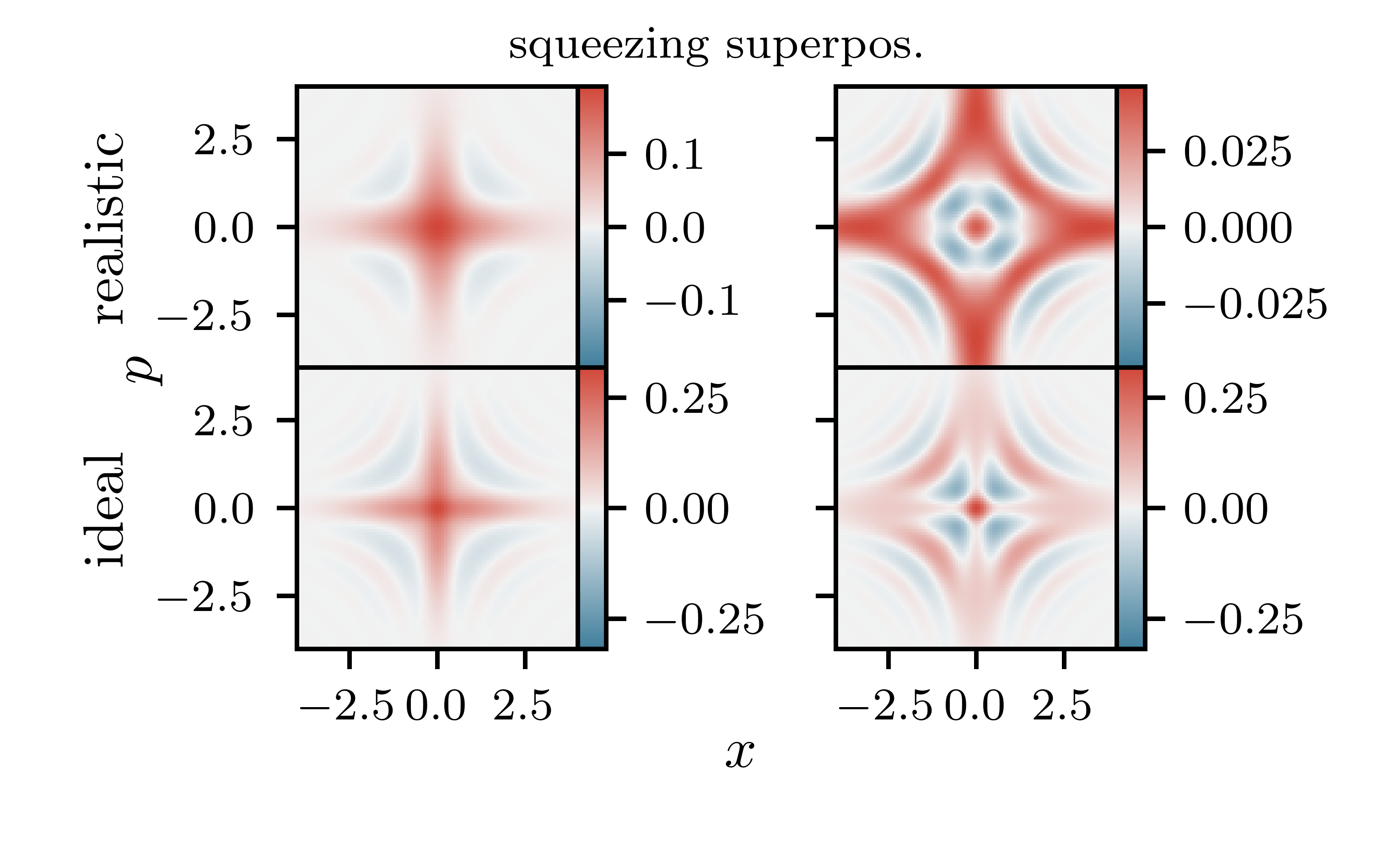}
    \caption{Wigner function for squeezing assuming either an ideal system or considering realistic system parameters (starting state $\bar{n}=0.1$ and heating rate of $\dot{\bar{n}} = 300\ \unit{q/s}$).}
\label{fig:supp_heated_vs_ideal}
\end{figure}

While we have good agreement between experiment and numerical simulation with realistic system parameters i.e.~finite initial oscillator occupation of $\bar{n}=0.1$ and non-negligible heating rate of $\dot{\bar{n}}= 300\ \unit{quanta/s}$, there is a difference compared to an ideal system with perfect initial groundstate and no heating rate.
The macroscopic properties, such as the overall shape of the Wigner function, stay the same, but certain details or features become more or less accentuated. We qualitatively compare the realistic and ideal states in Fig.~\ref{fig:supp_heated_vs_ideal}. The Wigner functions with finite temperature and heating effects are identical to the plots shown for Fig.~\ref{fig:equal_superpositions}HI.

We extract the WLN described in the previous section for the numerically simulated and experimentally reconstructed states.
The WLN for the experimental states is close to the value of the realistic simulation but reduced compared to the ideal states due to experimental imperfections. We present the extracted values in Table~\ref{tab:supp_wigner_negativities}. However, we emphasise that these values should give a qualitative estimation but do not represent a quantitative characterisation. 
We would need to improve the Wigner function reconstruction protocol to deal with spurious ripples in the Wigner function for larger $x,\, p$ values, which arise due to shot noise and other experimental imperfections during the tomography. We have windowed the Wigner function around the origin to avoid biasing the results. The window size is set such that $(\int_{\rm window} W(x,p) dx dp)/(\int W(x,p) dx dp) =0.95$. For reference, we also extract the minimum value of the Wigner function ${\rm min}(W)$, which is occasionally stated in the literature. However, we stress that this measure is bounded to ${\rm min}(W)\geq -1/\pi$ and only confirms that the state has Wigner negativity but does not measure resourcefulness. For example, the ideal Fock state $\ket{1}$ has a minimum value of ${\rm min}(W) = -1/\pi$.
\begin{table}[h!]
\centering
\begin{ruledtabular}
\begin{tabular}{ccc}
 Squeezed superposition & WLN & min($W$) \\ 
\hline
Even: exp &0.14 & -0.025\\
Even: realistic & 0.132 & -0.022 \\ 
Even: ideal & 0.408 & -0.050 \\ 
Odd: exp &0.209 & -0.025\\
Odd: realistic & 0.193 & -0.025\\ 
Odd: ideal & 0.756 & -0.177\\ 
\end{tabular}
\end{ruledtabular}
\caption{Extracted Wigner logarithmic negativities (WLN) and minimum negative value of the Wigner function min($W$) reached for the squeezed superposition. We compare the experimental reconstructed, realistic simulation and ideal simulation for both even and odd superposition states.}
\label{tab:supp_wigner_negativities}
\end{table}

\subsection{Reconstruction of the Wigner function}
\begin{figure}[ht]
\includegraphics[width=\linewidth]{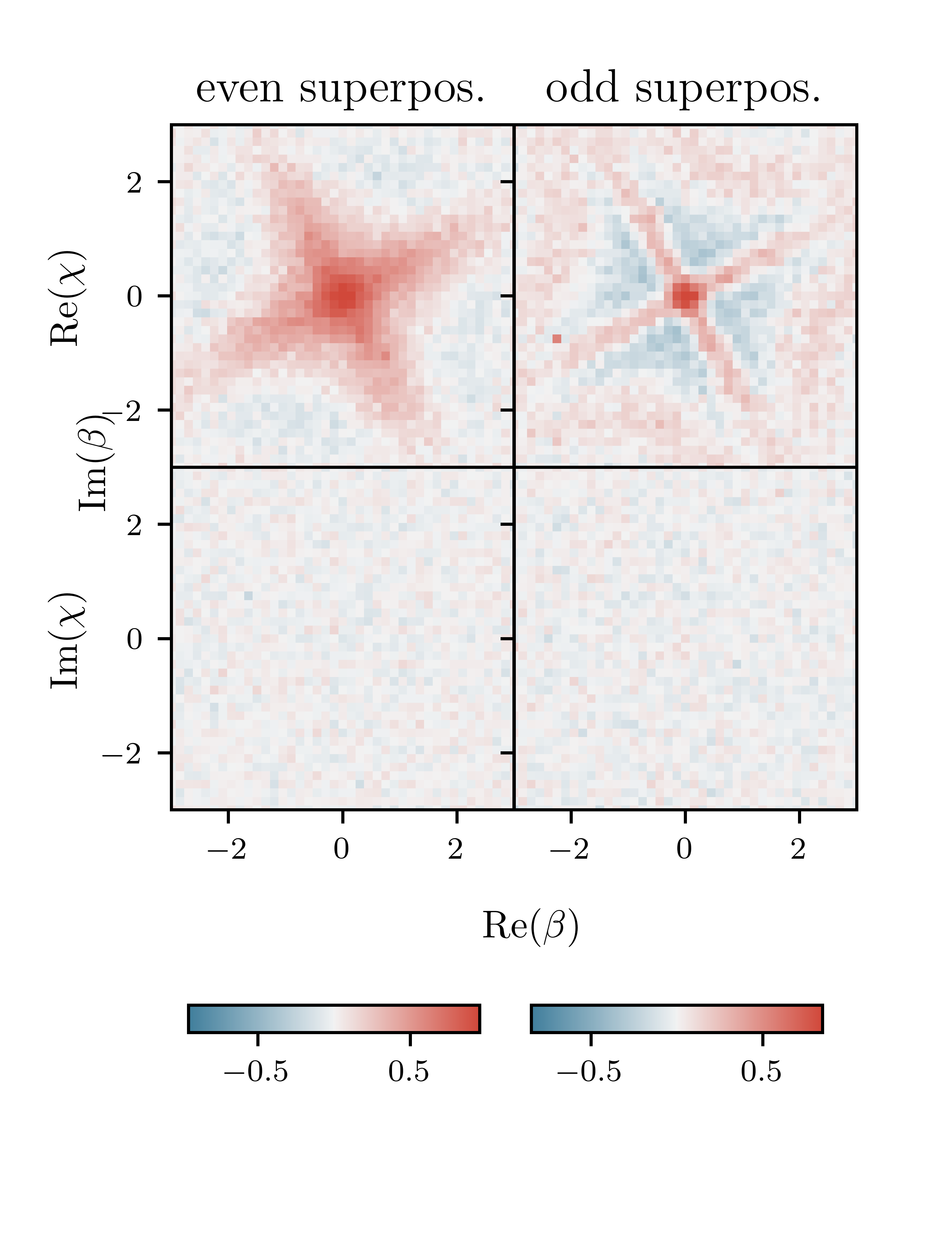}
    \caption{Characteristic function of Fig.~\ref{fig:equal_superpositions}B and Fig.~\ref{fig:equal_superpositions}C. The imaginary part ${\rm Im}(\chi(\beta))$ vanishes despite the Wigner functions complex pattern.}
\label{fig:supp_charact_fnc}
\end{figure}

We obtain the Wigner function via Fourier transform of the characteristic function $\chi(\beta) \in \mathbb{C}$, following Refs.~\citenum{fluhmann2020direct, bazavan2024squeezing}.
As an example, we provide the characteristic function of the even (Fig.~\ref{fig:equal_superpositions}B) and odd squeezed superposition states (Fig.~\ref{fig:equal_superpositions}C).
In Fig.~\ref{fig:supp_charact_fnc}, we show the experimentally measured characteristic function.
In the experimentally measured characteristic function, and consequently, in the reconstructed Wigner function, we find a tilt relative to the $x$, $p$ axes of $\sim  \SI{0.44}{\radian}$, with daily variations on the order of $\SI{0.06}{\radian}$. This tilt is the result of a constant motional phase offset between the SDF used for reconstruction and the axis of the squeezing interaction. To make the comparison to the numerically simulated data easier, we correct the shift by digitally rotating the reconstructed Wigner function by this amount.

\subsection{Density matrices of equal superpositions in Fig.~\ref{fig:equal_superpositions}}
\begin{figure}[ht]
\includegraphics[width=\linewidth]{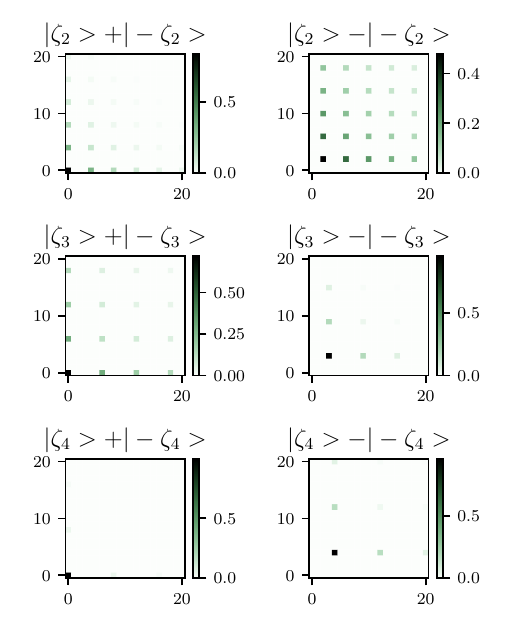}
\caption{Visual representation of the first 20 x 20 entries of the density matrices expressed in the Fock state basis for squeezed, trisqueezed and quadsqueezed superpositions shown in Fig.~\ref{fig:equal_superpositions} assuming a perfect ground state $\ket{0_{osc}}$ i.e.~$\bar{n}=0.0$ and no decoherence effects. We assumed squeezing magnitudes $\zeta_2 = \{1.12(5), 1.67(7)\}$, $\zeta_3= \{0.74(5), 0.74(5)\}$ and $\zeta_4=\{0.059(5), 0.16(1)\}$ as in Fig.~\ref{fig:equal_superpositions}.}
\label{fig:supp_density_matrices}
\end{figure}

To provide more intuition about the odd and even superposition states, we show the density matrices in the Fock basis (see Fig.~\ref{fig:supp_density_matrices}). As expected only the Fock states $\ket{k\cdot 2n}$ for the even superpositions and $\ket{k\cdot (2n + 1)}$ for the odd superpositions, where $n\in \mathbb{N}_0$ and $k$ the order of the nonlinear interaction, are populated.

\label{suppl_material}
\end{document}